# Panopticon[*]: a telescope for our times

Will Saunders,[1] , Timothy Chin, Michael Goodwin
Australian Astronomical Optics, Macquarie University, NSW, Australia

## ABSTRACT

We present a design for a wide-field spectroscopic telescope. The only large powered mirror is spherical, the resulting spherical aberration is corrected for each target separately, giving exceptional image quality. The telescope is a transit design, but still allows all-sky coverage. Three simultaneous modes are proposed: (a) natural seeing multi-object spectroscopy with 12m aperture over 3° FoV with ~25,000 targets; (b) multi-object AO with 12m aperture over 3° FoV with ~100 AO-corrected Integral Field Units each with 4" FoV; (c) ground layer AO-corrected integral field spectroscopy with 15m aperture and 13' FoV.

Such a telescope would be uniquely powerful for large-area follow-up of imaging surveys; in each mode, the AOmega and survey speed exceed all existing facilities combined. The expected cost of this design is relatively modest, much closer to $500M than $1000M.

## 1. INTRODUCTION

To achieve funding in these times, any major new telescope must provide a dramatic science gain for reasonable cost. Wide-field MOS spectroscopy is one area where large gains are feasible and very desirable, with several proposed new telescopes (WST [1], MSE [2], EST [3], Megamapper [4]). Another area where transformative gains look feasible is wide-field multi-object AO (MOAO) spectroscopy, where there are no funded projects with fields greater than ~14' [5,6].. Finally, the success of VIRUS [7] and MUSE [8] have shown the power of large area integral field spectroscopy (IFS), and large gains are possible here also.

We present a telescope design allowing all these three modes (MOS, MOAO, GLAO+IFS) simultaneously. The design is an evolution of WHAT? [9], also inspired by HET [10], SALT [11], and LAMOST [12]; and by the work of Lyndon-Bell [13] and Burge and Angel [14]. It has greatly reduced costs compared with a traditional fully-steerable telescope, having all large optics segmented and spherical or flat, and having no top end. The design is a transit telescope like LAMOST, but full-sky coverage is possible by switching the position of the rail-mounted primary. The design is a 'Poor man's Schmidt', with spherical primary but without a corrector at the entrance pupil. The spherical aberration is corrected at the focus, like HET or SALT. Spherical aberration varies as the inverse cube of the telescope speed; for an f/5 primary, and small (~1") FoV, the size of the spherical aberration corrector (SAC) can be reduced from the oil-drum size of SALT to a few mm, small enough to put into a fiber positioner or an autonomous deployable front-end. If this is done with two mirrors (like SALT), one of the mirrors can be deformable, allowing MOAO use. The resulting telescope image quality is uniformly excellent.

The nominal proposed aperture is 12m in wide-field use, and 15m in IFS use, but nothing in the design precludes a larger or smaller size. Because the primary has no axis, FoV is limited only by vignetting considerations; a 3º design is presented, but a 2.5º FoV would very significantly reduce the vignetting losses. The focal surface is near ground level, fixed (except for image rotation), 3.15m across and horizontal. This means that very ambitious instrumentation can be accommodated nearby and underground, very helpful for passive thermal control, and with very short (~10m) fiber lengths.

---

[*] The original Panopticon was an 18th century model prison allowing permanent 360º surveillance.

[1] will.saunders@mq.edu.au

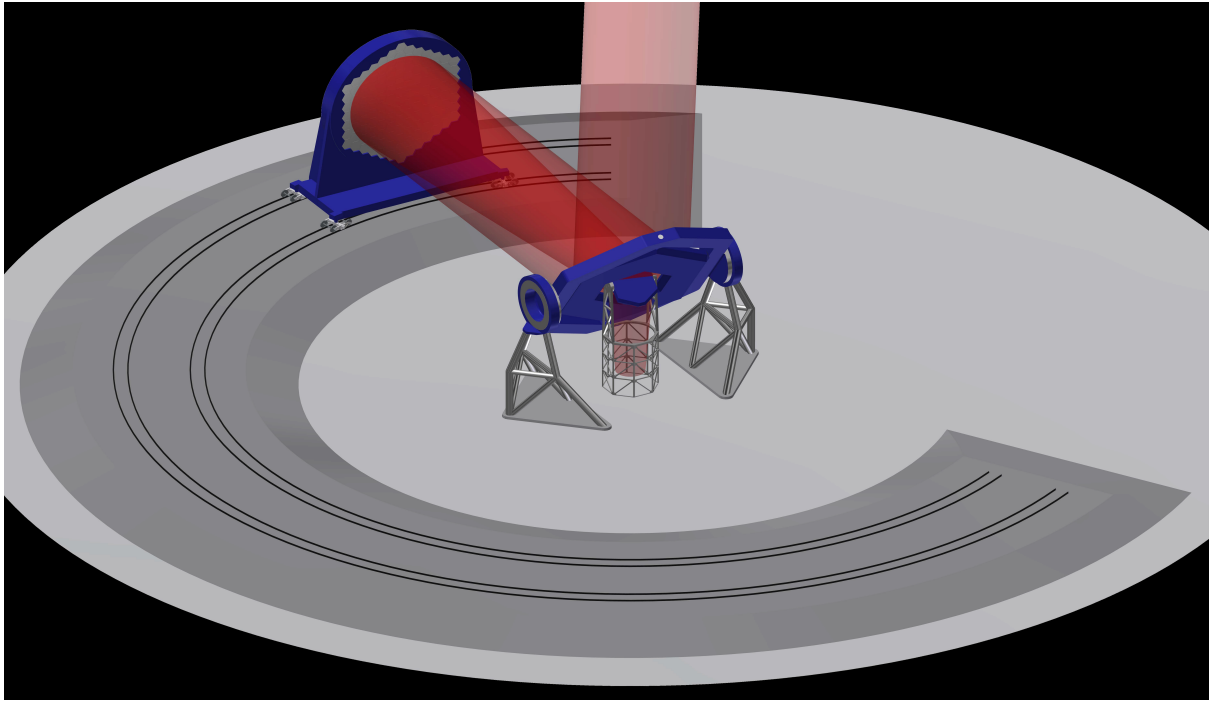

**Figure 1. Physical layout showing flat steering mirror M0, spherical primary M1, fold mirror M2 at center of M0, and horizontal focus near ground level. M0 can move freely in elevation north and south, to ZD~65º in each direction. It can also tilt laterally by ±11.5º, giving 3 hours tracking for all fields. M1 is rail-mounted and can be moved between northern and southern stations to switch the telescope orientation.**

A large IFS has been incorporated, as envisaged for WST. The SAC for IFS use consists of a single fixed concave 4.0m mirror, the re-imaged focus is 35m away from the primary focus, giving ample room for the massive banks of spectrographs required without encroaching on the MOS and MOAO spectrographs.

## 2. DESIGN OVERVIEW

The spherical primary (M1) is 16.1m tip-to-tip with 60m focal length, and is fixed with axis horizontal. It is fed by a flat gimbal-mounted steering mirror (M0) 48m away. There is a fold mirror M2 at the center of M0, which also allows tip-tilt correction for wind-shake. The focal surface is below the steering mirror close to ground level, it is hexagonal and horizontal. There is a SAC for each target, which always defines the stop. The SAC consists of a mini-lens for MOS mode, a pair of mirrors for MOAO mode and a single mirror in IFS modes. These mirrors allow AO image correction with no additional surfaces.

The fixed and large (3.15m) focal surface allows a huge MOS positioner, while still allowing a very large (13') IFS at the field center. A transparent field-plate, segmented as necessary, allows large numbers of autonomous MOAO units to deploy on the upper surface, at the cost of vignetting some fraction of the MOS and/or IFS fields.

The telescope is a transit design; the steering mirror can be tilted to any elevation >25º north or south, but has just ±11.5º tilt in the transverse direction. Each field can be observed for 3 hours as it traverses the meridian (allowing tracking through or even behind zenith). On any given night, the telescope is either north-facing or south-facing. However, the primary is on rails and the orientation can be completely reversed between north and south, so that overall sky coverage is as good as any fully steerable telescope.

The three mirrors are all segmented, either flat or spherical, and assume the ELT segment size. The primary mirror has 126 segments (the central segment is missing), the steering mirror is the same width but 1 segment (1.2m) longer, with a 26-segment central hole and 116 segments. There is marginal benefit from being phased, but this is not essential.

Significant vignetting is accepted at the edges of the 3° field, especially at small ZDs. Figure 5 shows the vignetting. Vignetting losses are particularly undesirable in a MOS telescope, since the S/N is reduced while spectrograph costs remain unaltered. A smaller FoV and faster primary speed would both reduce the vignetting losses significantly. In IFS mode, the vignetting is 22% at ZD=65° and 39% at zenith.

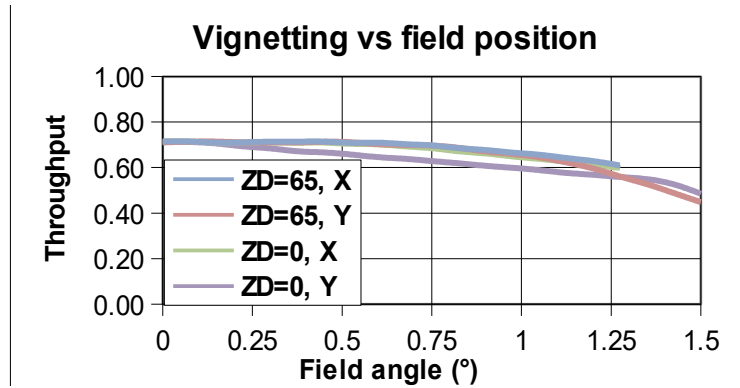

Figure 2. Vignetting as a function of field angle.

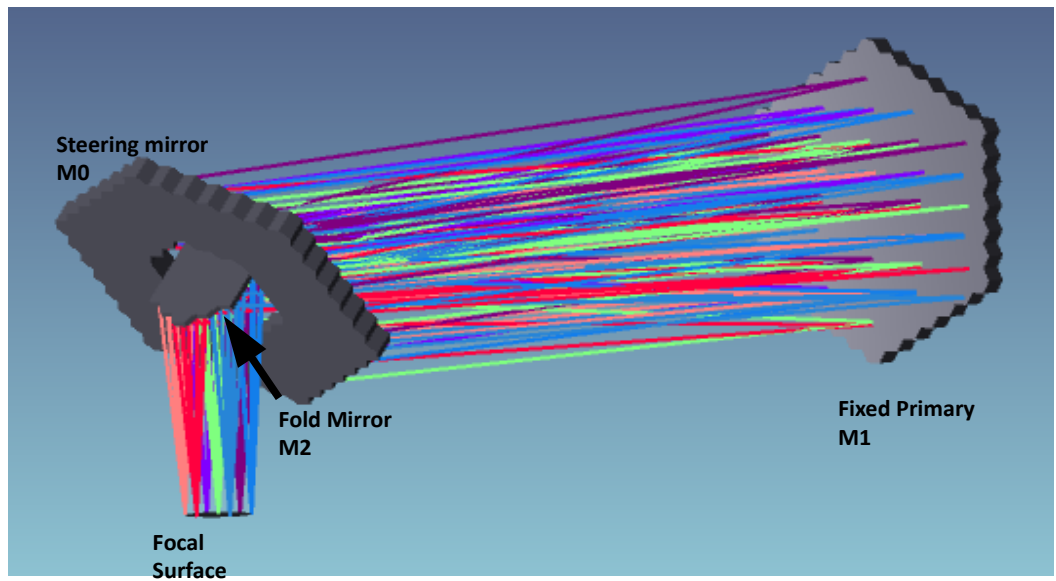

Figure 3. Optical layout. The primary is the only large powered surface.

No baffling is included in the design, since the stop is always well defined by the SAC for each mode. However, for thermal NIR use, the useful FoV will be limited by the vignetting at M0 and M1 (since this allows warm surfaces into the beam). At ZD=20°, the unvignetted field is ~1° across and hexagonal; the field is elongated at larger ZDs and foreshortened at smaller ZDs.

## 3. MOS MODE

The simplest mode is natural seeing MOS. The proposed positioners are the Potsdam 'FLEX' design [15]. ~25,000 actuators could be accommodated while keeping a generous pitch, ~20mm. The SAC is a doublet lens 23.5mm long and just 4.1mm across, mounted in the end of the positioner, and delivering an f/3 image with $d_{80}$ < 0.1". The exclusion radius round each target is 14".

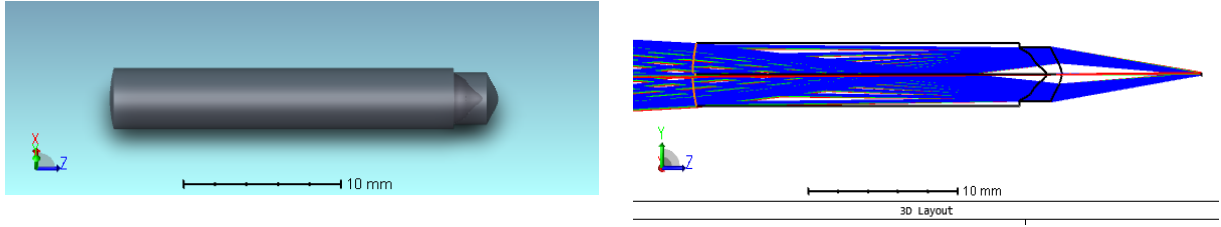

**Figure 4.** SAC corrector doublet for MOS mode, 23.5mm long and 4.1mm (14") across. The glasses are S-FPL51Y and PBL6Y. The cutout on the right is to allow mounting in a fiber positioner while minimizing the overall diameter.

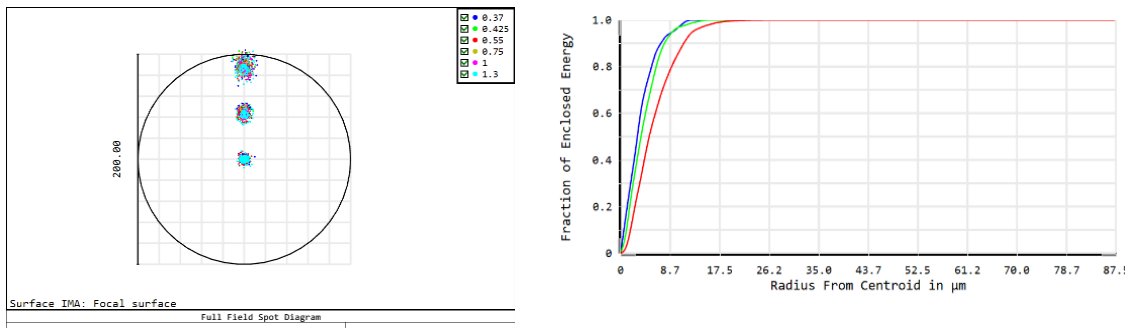

**Figure 5. Left:** Spot diagram for MOS mode at zenith. Fields are at 0, 0.25", 0.5", circle is 1". **Right:** Geometric encircled energy for the same fields, maximum radius is 0.5", $d_{80}$ < 0.1".

The image is projected onto a mini-hexabundle of 7 fused fibers, each of ~0.4" diameter. Dissecting the image like this offers multiple powerful advantages: (a) differential atmospheric refraction can be corrected for in the extraction, (b) it allows reasonable spectrograph camera speeds, (c) it gives some spatial information for resolved targets, (d) it gives a large S/N gain for non-circular resolved targets, and (e) it gives a dramatic increase in survey speed in the best conditions (since the fiber co-addition will be S/N-weighted). Figure 6 shows the dependence of survey speed on seeing for a subdivided vs monolithic fiber aperture of the same area. When this speed is integrated over the Paranal seeing profile (assumed lognormal with median 0.66" and $\sigma$=1.35), and neglecting all overheads, the subdivided fiber gives essentially the same overall survey speed as the single fiber. However, for more difficult targets observed only in good conditions, the hexabundle offers up to a magnitude increase in sensitivity.

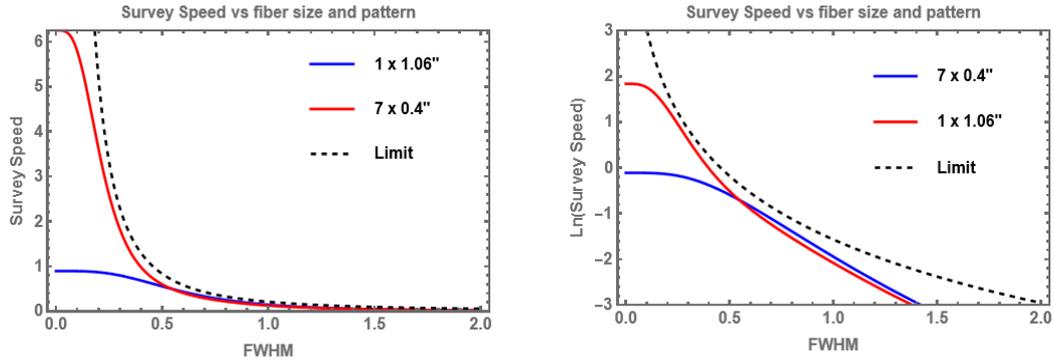

**Figure 6. Relative survey speed vs delivered image FWHM for a 7-fiber hexabundle (1:1.1 core/clad ratio) and a single fiber of the same area, (a) linear and (b) log plot. The dashed line shows the theoretical limit (the S/N content of the 2D image). Moffat $\beta$=3 profiles are assumed, and read-noise, positioning errors and overheads are neglected for all curves. Note (a) the strong dependence of survey speed on image quality even for a single fiber, (b) the dramatic further increase in speed offered by a hexabundle in the very best seeing, and (c) the modest hexabundle speed penalty in poor seeing.**

There is no ADC in MOS mode (all ideas gratefully received). However, the overall penalty to survey speed is modest for most programs, with the overall diameter of the hexabundle (1.28") equalling the atmospheric dispersion between 400nm and 1000nm at zenith distance 45º at 2500m altitude. Only programs requiring S/N-critical coverage at both ends of the spectrum and observed at large ZDs would be significantly impacted, and even in these cases each field could simply be observed twice, with 'blue' and 'red' pointings (the latter could be in bright time), and the spectra co-added.

The most undesirable feature of this design is that the tolerance on the mini-lens angular alignment is extremely tight, ~2'. This rules out 'tilting spine'-type positioners, and may be very difficult for other positioner designs. More tolerant mini-lens designs are being investigated.

## 4. IFS MODE

IFS mode picks off the central 13' of the focal surface. To minimize the loss of MOS field, the light first passes through a fused silica cone the same depth as the MOS positioners, restricting the loss of MOS field to ~20'. There is a doublet (BSL7Y/PBL6Y) field lens. The SAC consists of a fixed concave 4m mirror, while a large fold mirror sends the beam to the basement spectrograph room. These mirrors are conjugate to 350m above M0, and midway between M1 and M2 respectively, ideal for GLAO image correction. The deformable mirrors (DM) are assumed to use the TNO amplified voice-coil technology [16], allowing a mirror of reasonable thickness and material. There is a K-mirror derotator, which could be anywhere in the optics train, but just before focus is convenient and compact. The K-mirrors can be shifted and tilted, allowing arbitrary sub-field selection. The final focus is 30m away from the telescope vertex, giving plenty of room for the IFS spectrographs without fouling the MOS or MOAO spectrographs. The IFS uses almost the full aperture of the telescope, 15m circular. The final focus is f/12, (or as needed otherwise) with a plate scale 877μm/" and $d_{80}$ < 0.1". The focal surface is pupil-centric and concave, with a RoC 330m.

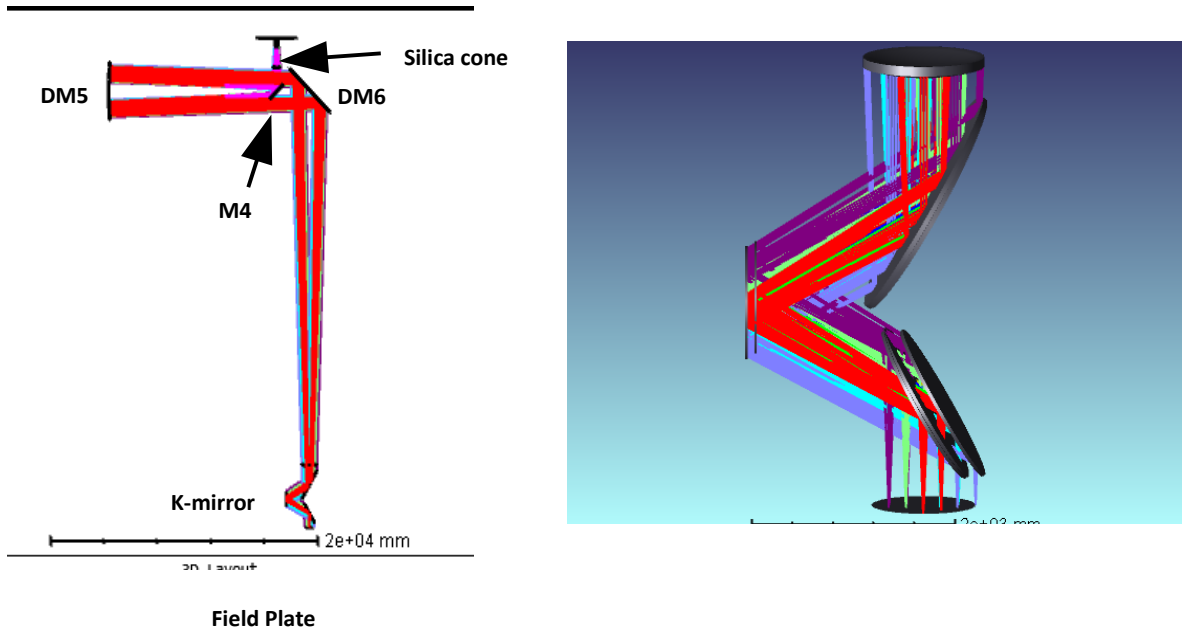

**Field Plate**

**Figure** 7. **(a) Overall** IFS layout. There is a fused silica cone at entry to reduce the loss of field for MOS mode. DM5 and DM6 are deformable, M4 is suitable for fast tip-tilt. (b) The mirrors of the K-mirror can be tilted and shifted as shown to allow selection of a sub-field of the whole 13' FoV.

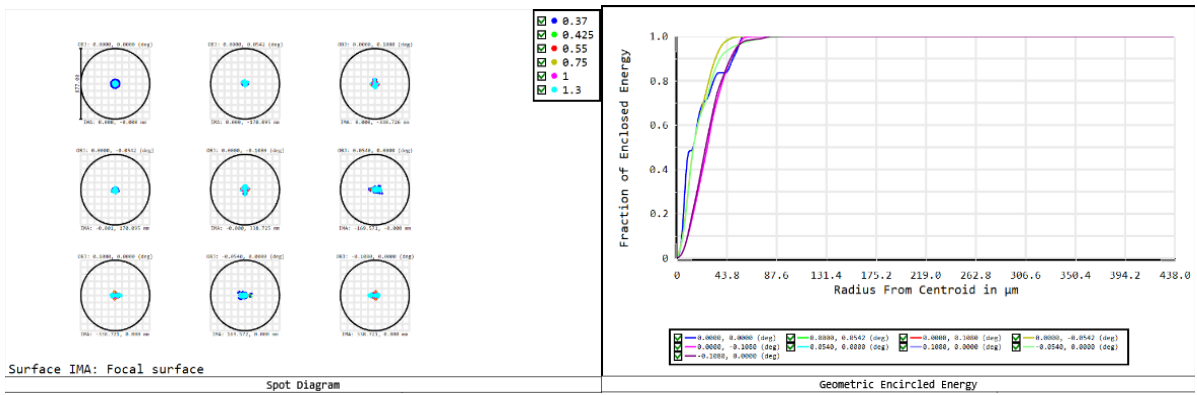

**Figure 8. (a) Spot diagram for IFS use at zenith, circle is 1" diameter. (b) and geometric encircled energy for IFS use, the plate scale is 877µm/" and $d_{80}$ < 0.1".**

## 5. MOAO MODE

MOAO mode require a laser guide star (LGS) and a natural guide star (NGS) for each target, since the targets will be much too far apart to share them. Each autonomous front end then needs a deformable mirror, a natural guide star imager and a wavefront sensor (WFS) for the LGS.

The DM is assumed to be a Kilo-C-3.5 from Boston MicroMachines [17], incorporated into the SAC. The design also incorporates a novel atmospheric dispersion corrector (ADC) design, with a wedged doublet and a wedged Mangin mirror providing the ADC action when counter-rotated. The image has 4" FoV at f/5, with $d_{80}$ < 0.1", feeding a 300-fiber hexabundle with 0.2" sampling.

The LGS light is picked off via a dichroic to a Shack-Hartmann sensor (so this is an open-loop system). The difference in LGS and science foci means that unscattered laser light cannot enter the science fibers, even if it is not reflected by the dichroic (it is this that limits the FoV to 4"). A second

NaD filter at fiber input ensures suppression of the LGS light into the science fibers by >30$^m$, rendering it negligible.

The LGS beams are launched from the interstices of the primary mirror segments and use the steering mirror to project towards the targets. The launch telescopes are fixed, except for ±0.5° tip/tilt adjustment. The beams can then pass within 1m of the entrance pupil center, reducing LGS elongation to <0.16". There are 108 suitable interstices, so 108 MOAO units are proposed, fed by 18 lasers. The TOPTICA lasers [18] are suitable, though their annual production is currently only ~2/year.

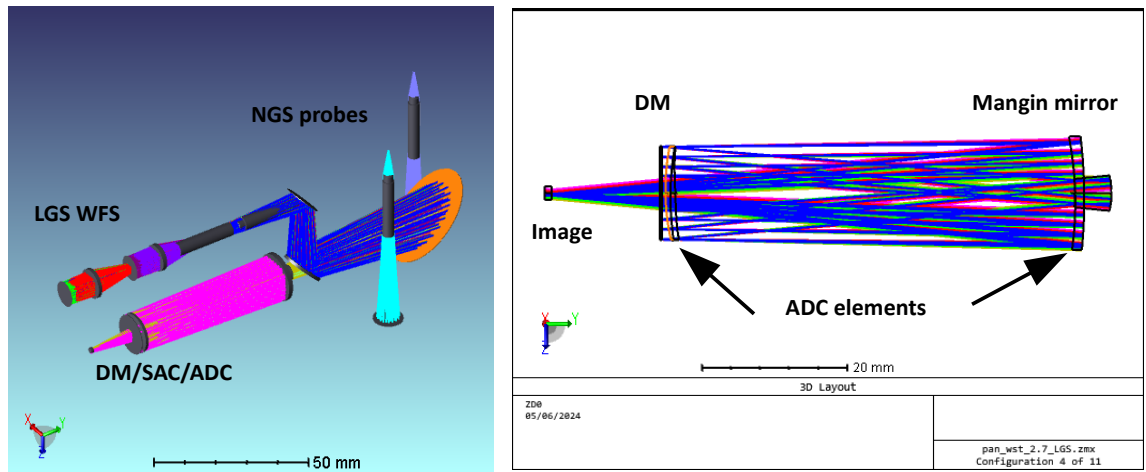

**Figure 9. Left: MOAO optics. The LGS WFS moves in focus as needed, limiting positions shown in red and purple. The NGS probes are autonomous starbug-mounted units with a few arcmin patrol radius; the images are transmitted to unused corners of the Shack-Hartmann imaging camera by short (~150mm) coherent bundles. Right: DM/SAC/ADC unit. The ADC elements (doublet at left, Mangin mirror at right) are wedged and tilted; counter-rotation then gives an ADC action.**

Two 'tethered' starbug-mounted [19] NGS units are proposed for each unit, with the same optics as the MOS front-ends, and with the light imaged onto short (~150mm) cohererent bundles, and re-imaged onto unused corners of the Shack-Hartmann imaging camera.

The overall size of the unit depends on the size of the camera and DM housing, but a footprint 150mm x 50mm looks reasonable, vignetting just 0.1% of the focal surface for each unit. The automation of the main unit would be via 3 starbugs and a 'hoverboard' [20].

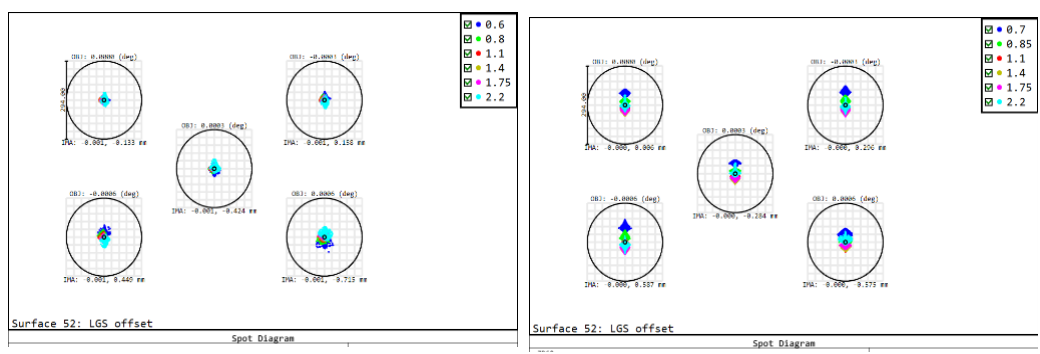

**Figure 10. Spot diagrams at (a) zenith and (b) at ZD=60º. Circle is 1".**

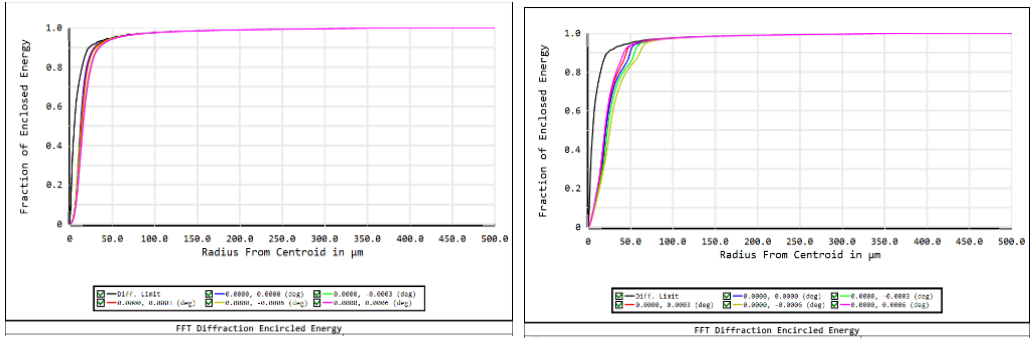

**Figure 11.** Encircled energy at (c) zenith and (d) ZD=60º. X-limit is ½", $d_{80}$ < 0.1" everywhere.

## 6. DELIVERED IMAGE QUALITY

In all modes, the optical design contributes negligibly to the delivered image quality (DIQ) budget. The simplicity of the optics means that 'as-built' errors should also be very small. The DIQ will then be overwhelmingly dominated by the seeing. The design has an inescapable ~80m of additional path length near ground level compared with a Cassegrain telescope, and this will impact the seeing. For MOAO and IFS modes, the AO will clean this up, since this ground-layer component is slowly changing and has large isoplanatic angle. For MOS mode, we can make a preliminary estimation of the penalty. Assuming a $C_n^2$ value at ground level at Paranal (the best studied Chilean site) of ~4 x 10$^{-16}$ m$^{-2/3}$ [21], gives a nominal contribution to the seeing of ~0.16" (to be added at the 5/3 power to the site seeing). However, this overestimates the effect – the distortions in each direction between M0 and M1 are highly anti-correlated (eg no tip-tilt). In any case, this is less than the *gain* in image quality offered by the telescope optics over any plausible prime focus or Cassegrain design. Clearly a detailed study is required, but this design can be expected to deliver a significantly *better* delivered image quality overall than a traditional telescope. As well as impacting survey speed directly, as shown in figure 6, this feeds directly into MOS spectrograph cost and risk via the reduced etendue and camera speeds.

## 7. COMPARISON WITH EXISTING TELESCOPES

The AOmega product of this telescope in MOS mode is an order of-magnitude greater than any existing facility, in fact greater than the combined AOmega of *all* spectroscopic facilities, existing or under construction. Table 1 shows a crude comparison. As well as the simple AOmega product, a generalized MOS survey metric is shown, defined as

$$P = A \Omega^\alpha N^{1-\alpha} / d^\beta$$

where $A$ is aperture area, $\Omega$ is FoV solid angle, $N$ is the number of fibers, $d$ is the median image quality. $\alpha$ reflects the relative importance of FoV and fiber numbers, $\alpha = ½$ is assumed here, i.e. equally important. The exponent $\beta$ depends on target type; $\beta=0$ is appropriate for bright stars, $\beta=2$ is appropriate for unresolved faint sources (eg QSOs). For marginally resolved faint sources (galaxies, Ly-$\alpha$ blobs) and also stars with comparable counts from star and sky, $\beta=1$ is an excellent approximation. Factors for efficiency, time fraction and spectral resolution elements are neglected. Survey speeds for bright stars and faint galaxies are shown, in units of 'SDSSs'.

| Facility | A/m$^2$ | $\Omega$/dg$^2$ | N | d(") | A$\Omega$ | $P_{bs}$ | $P_{fg}$ |
|---|---|---|---|---|---|---|---|
| SDSS | 3.68 | 7.06 | 600 | 1.4 | 26.0 | 1 | 1 |
| 2dF | 10.0 | 3.14 | 400 | 2.0 | 31.4 | 1.5 | 1 |
| LAMOST | 14.5 | 19.6 | 4000 | 3.5 | 2842 | 17 | 6.8 |
| DESI | 9.5 | 8.04 | 5000 | 1.5 | 76.4 | 8 | 7.5 |
| PFS | 50 | 1.33 | 2400 | 0.6 | 66.5 | 12 | 28 |
| 4MOST | 12 | 4.90 | 2436 | 0.75 | 58.8 | 5.5 | 10 |
| MegaMapper | 28 | 7.06 | 20000 | 0.65 | 198 | 44 | 95 |
| MSE | 73 | 1.50 | 18720 | 0.6 | 110 | 51 | 120 |
| LSSTspec | 35 | 9.60 | 8640 | 0.65 | 339 | 42 | 90 |
| WST | 89 | 3.14 | 22000 | 0.65 | 432 | 96 | 210 |
| Panopticon | 81 | 5.85 | 25000 | 0.65 | 475 | 130 | 280 |

**Table 1. Nominal survey speeds for selected survey facilities, existing, under construction, and proposed. Partially taken from [4].**

## 8. SITING, SEEING, COST, VARIANTS

The design is speculative, though guided by the requirements for WST. The lack of an ADC in MOS mode means that a high site is preferred. Good seeing has an enormous impact on both survey speed and spectrograph costs, and this design can take full advantage of the best seeing. For both reasons, somewhere on or near the Chajnantor plateau seems the obvious place to look for a suitable site, but many other sites would be suitable.

Estimated costs are equally speculative. But the SALT telescope has mirror area 2/3 as large as each of the two main mirrors here, and was built for $20m [22]. LGS lasers, MOAO DMs, IFS DMs, MOS positioner and IFS image slicers are all estimated at one-few tens of $M. A telescope cost of $300M seems plausible, not including the spectrographs.

If the primary mirror was f/4 instead of f/5, it would be possible reduce the M0-M1 distance, raise the focal surface, and to greatly reduce the central obstruction. The largest penalty is an increase in the closest MOS target separation to 6mm or 27". Vignetting losses could also be greatly reduced is a 2.5º field was accepted. As well reducing the required size of the hole in M0 from 24 to 10 segments, the vignetting at the edges of the field is greatly reduced also.